\documentclass[aps,prl,reprint,twocolumn,superscriptaddress,amsmath,amssymb,showpacs]{revtex4-1}
\usepackage{amsmath}
\usepackage{amsbsy}
\usepackage{natbib}
\usepackage{graphicx}
\usepackage{epstopdf}
\usepackage{color}
\usepackage{times}
\usepackage[colorlinks,bookmarks=false,citecolor=blue,linkcolor=red,urlcolor=blue]{hyperref}

\def\be {\begin{equation}}
\def\ee {\end{equation}}

\begin{document}
\title{Unveiling the Nature of Three Dimensional Orbital Ordering Transitions:\\ The Case of $e_g$ and $t_{2g}$ Models on the Cubic Lattice}
\author{Sandro Wenzel}
\email{sandro.wenzel@epfl.ch}
\affiliation{Institute of Theoretical Physics, \'Ecole Polytechnique F\'ed\'erale de Lausanne (EPFL), CH-1015 Lausanne, Switzerland}
\author{Andreas M. L\"auchli}
\email{laeuchli@comp-phys.org}
\affiliation{Max-Planck-Institut f\"ur Physik komplexer Systeme, N\"othnitzer Str.\ 38, D-01187 Dresden, Germany}

\begin{abstract}
We perform large scale finite-temperature Monte Carlo simulations of the classical $e_g$ and $t_{2g}$ orbital models on the simple cubic
lattice in three dimensions. The $e_g$ model displays a continuous phase transition to an orbitally ordered phase.
While the correlation length exponent $\nu\approx0.66(1)$ is close to the 3D XY value, the exponent $\eta \approx 0.15(1)$ differs
substantially from O(N) values. At $T_c$ a $U(1)$ symmetry emerges, which persists for $T<T_c$ below a crossover length scaling as $\Lambda \sim \xi^a$,
with an unusually small $a\approx1.3$. Finally, for the $t_{2g}$ model we find a {\em first order} transition into a low-temperature lattice-nematic phase without orbital order.
\end{abstract}
\date{\today}
\maketitle
Orbital-only models emerged recently as prototype systems enabling the
understanding of relevant aspects of the collective dynamics of
orbital degrees of freedom~\cite{vandenBrink_NJP}. In a different
context, orbital-like models are attracting considerable theoretical
interest due to their ability to sustain topologically ordered phases
with possibly anyonic excitations, as exemplified by the Kitaev
honeycomb model~\cite{kitaev-model}.  In a similar spirit the orbital
compass model~\cite{kugel82} can serve as a basic model to understand
topologically protected Josephson junction
qubits~\cite{doucot:024505}, which have recently been realized
experimentally~\cite{gladchenko-2008}.

A variety of properties have already been uncovered for orbital-only
models, but most of these are restricted to ground state or
low-temperature properties. Much less is known about
finite-temperature properties and in particular the nature of thermal
phase transitions. Those might display new critical phenomena, as a
common feature of all these systems is a manifest coupling between
order parameter space and real space, which distinguishes them from
the well studied O(N) (such as Ising, XY and Heisenberg)
models~\cite{vicarireview}.

In this Letter, we present a comprehensive Monte Carlo (MC)
investigation of the nature of the finite-temperature phase
transitions in two popular orbital-only models on the
three-dimensional (3D) cubic lattice: the $e_g$ and the $t_{2g}$
models~\cite{vandenBrink_NJP}. We study here the classical versions
because the corresponding quantum models have a sign problem
precluding Quantum Monte Carlo approaches, and because in
Ginzburg-Landau theory one typically expects quantum and classical
versions of a same model to have the same critical properties,
although exceptions are possible.  The $e_g$ and the $t_{2g}$ models
are also often called the $120^\circ$ and compass models,
respectively. While the thermal phase transition in the
two-dimensional (2D) compass model has been the focus of recent
studies~\cite{mishra:207201,wenzelQCMPRB,wenzelCM2010} and clarified
to belong to the 2D Ising universality class, little is known about
the $e_g$ and $t_{2g}$ models in 3D - although potentially of more
direct relevance for the description of collective orbital
phenomena~\cite{vandenBrink_NJP}. We start by discussing the $e_g$ model
and its critical properties in some detail and turn then briefly to the $t_{2g}$ model
towards the end of this paper.

\paragraph{The $e_g$ model --- }
The $e_g$ model (EgM) is defined by the Hamiltonian~\cite{vandenBrink_NJP}
\begin{equation}
    \label{eqn:eg_model}
    \mathcal{H}_{e_g}= -\ J\sum_{i, \alpha } {\boldsymbol{\tau}}_i^\alpha {\boldsymbol{\tau}}_{i+\mathbf{e}_\alpha}^\alpha,
  \end{equation}
  where $\boldsymbol{\tau}_i$ is an auxiliary three component vector obtained by
  an embedding of the orbital degree of freedom $\mathbf{T}_i=(T^z_i,T^x_i) \in S^1$:
  \be
  {\boldsymbol{\tau}}_i = \begin{pmatrix} -1/2 & \sqrt{3}/2 \\ -1/2 & -\sqrt{3}/2  \\ 1 & 0 \end{pmatrix} \mathbf{T}_i.
  \ee
  The $\mathbf{e}_\alpha$ denote the positive unit vectors in the $\alpha \in \{x,y,z\}$  cartesian directions. Note that the coupling in $\boldsymbol{\tau}$-space depends on the 
  spatial orientation of the bond. The coupling constant $J$ is set to one in the following, corresponding to ferromagnetic interactions. Note that results for antiferromagnetic interactions 
  can be deduced from results using ferromagnetic couplings~\cite{Rynbach}.
  
  The classical EgM (\ref{eqn:eg_model}) has a sub-extensive ground state degeneracy 
  which is lifted at finite temperature by an order by disorder
  mechanism~\cite{nussinov-2004-6}, leading to six
  discrete ordering directions $\mathbb{T}^o_n = \left(
  \cos[n\ 2\pi/6], \sin[n\ 2\pi/6] \right)$ with $n=0,\dots,5$. This
  analytical prediction has been verified using classical MC
  simulations~\cite{tanaka:267204,Rynbach}, and at higher temperatures
  a continuous phase transition to a disordered phase has been found.
  The prominent question of the universality class of the
  finite-temperature phase transition is however still open, both
  analytically and numerically. 
  Comparing to related systems with a similar low-temperature phase,
     several different scenarios
    seem possible: i) a continuous transition in the
    universality class of the 3D XY model, as e.g.~in the $Z_6$-perturbed
    XY models~\cite{Jose_PRB,Blankschtein_PRB,Oshikawa_Zn,Louclock},
    ii) distinct universality classes, as reported in classical dimer
    models on the cubic lattice~\cite{FAlet-dimer,CharrierAlet2010} or
    iii) a first order transition, as in a six-state ferromagnetic Potts model 
    in 3D, or a Heisenberg ferromagnet with a specific cubic
    anisotropy~\cite{vicarireview}.
    In the following, we shall resolve this fundamental question
    and answer which scenario is realized for the $e_g$ and $t_{2g}$ models.
%
\begin{figure}
\includegraphics[width=0.495\linewidth]{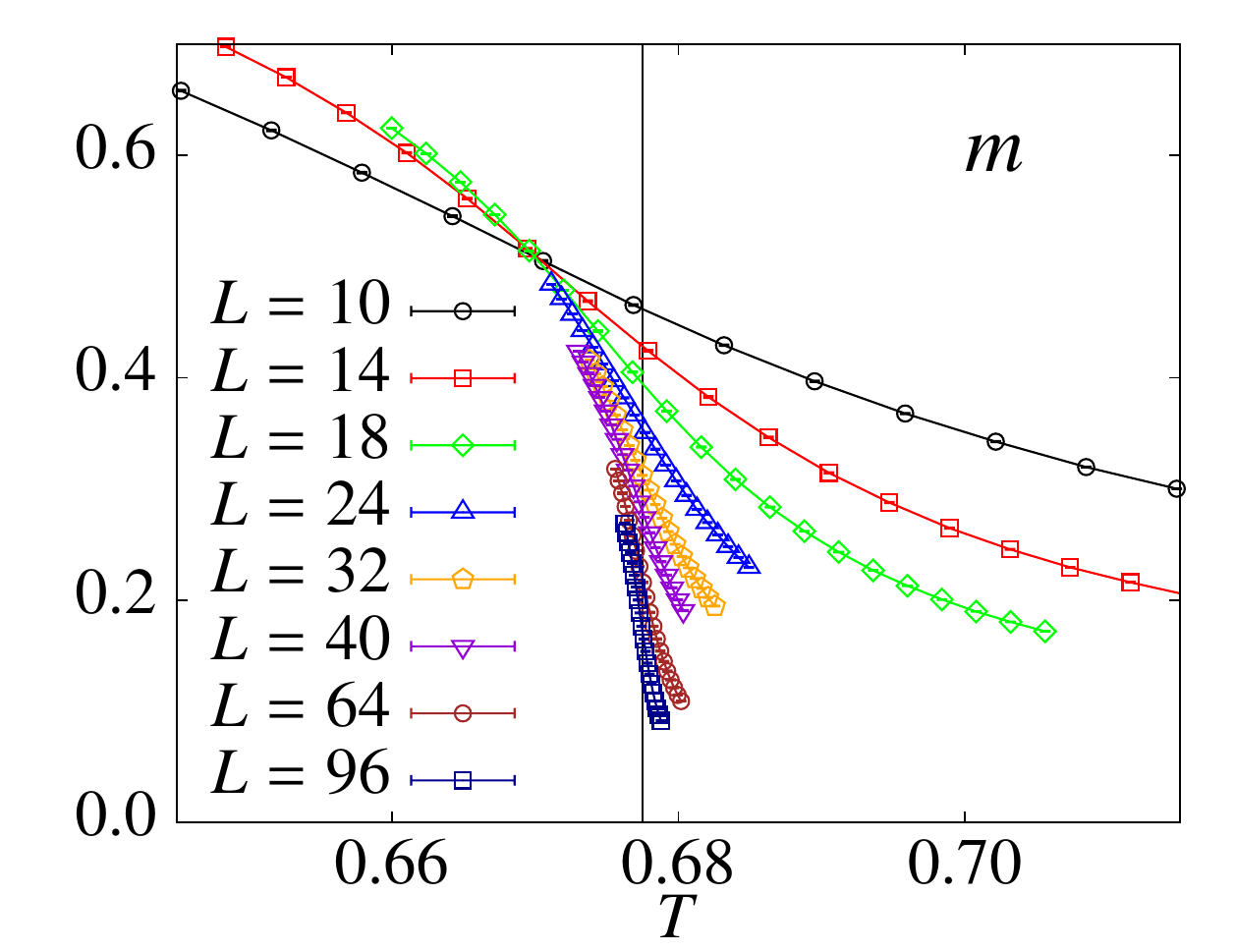}
\includegraphics[width=0.495\linewidth]{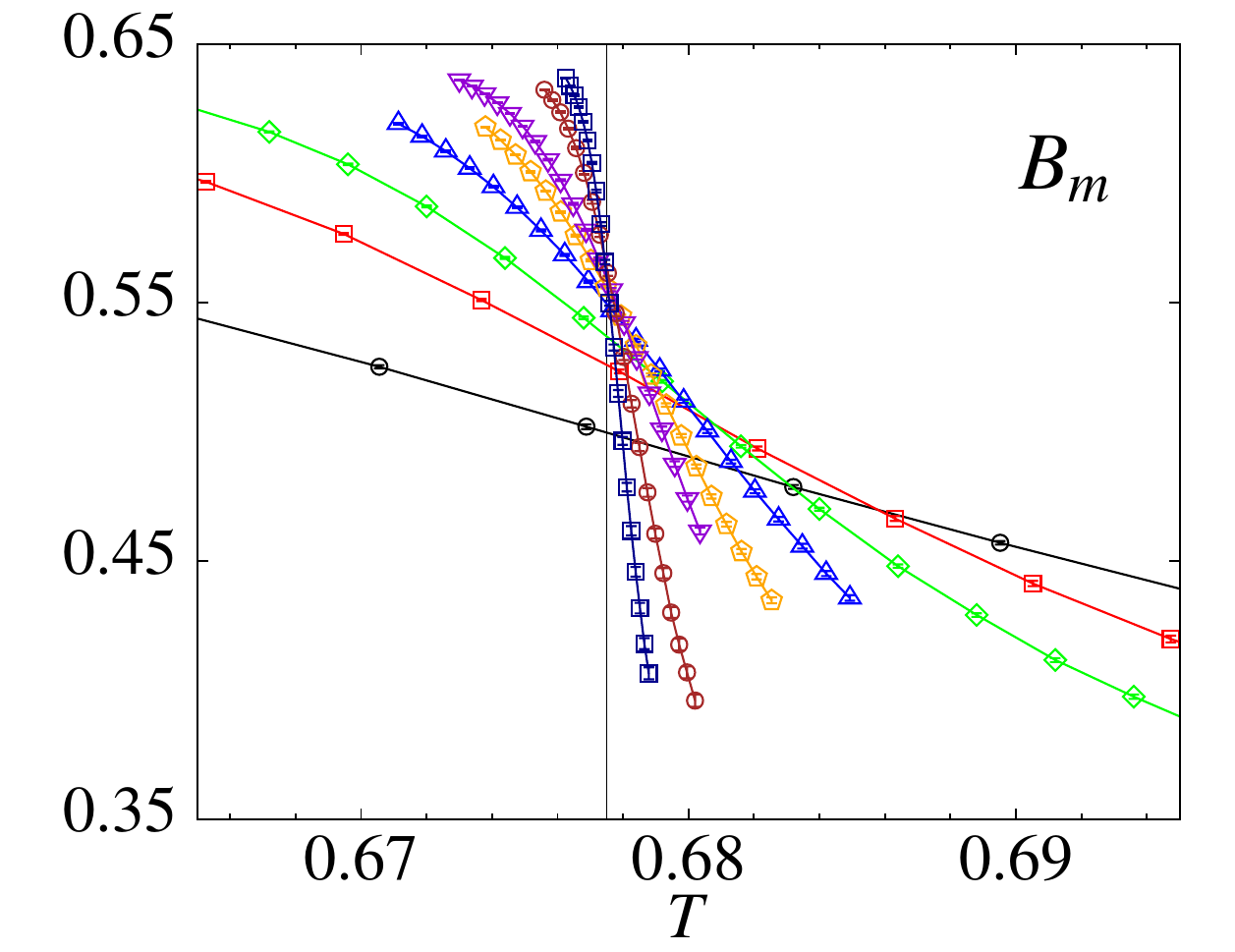}
\caption{(Color online). {\em $e_g$ model}: The order parameter $m$ (left plot) and the associated Binder cumulant $B_m$ (right plot) as a function of temperature 
$T$ for different linear system sizes $L$. The vertical line indicates the location of the critical temperature.}
\label{fig:overview}
\end{figure}
%
%
\paragraph{Simulation technique and observables ---} We consider
the classical Hamiltonian \eqref{eqn:eg_model} on a simple cubic lattice of side length $L$ and volume $N=L^3$ and perform
state-of-the-art MC simulations along the lines of Refs.~\cite{wenzelQCMPRB,wenzelCM2010}. Simulations were performed for
lattice sizes $L=8,\dots,96$. To obtain the reported accuracy, we collected
$10^6$ and more independent MC measurements per data point. MC runs using periodic boundary conditions (PBC) show clear signals of a transition
to an ordered phase in accordance with Ref.~\cite{Rynbach}. However, as further demonstrated 
below, we find that there are severe finite-size corrections using PBC.
Fortunately, we possess an efficient tool to substantially reduce the strong finite-size effects of PBC by employing  screw-periodic boundary
conditions (SBC), as shown recently for the 2D compass model in Ref.~\cite{wenzelCM2010}. Here, we shift the cube $L/2$ steps in the $x$-direction 
when leaving the $zy$-face (plus cyclic permutations), which we empirically find 
to minimize finite-size effects~\cite{WenzelEGlong}.  A natural order parameter to detect orbital ordering in the following is:
\be
m=(1/N) \sqrt{( \sum_i T^z_i )^2 + ( \sum_i T^x_i)^2},
\label{eqn:m}
\ee
while the complementary quantity $D$ indicates a directional ordering of the bond energies:
\be
D=(1/N)\sqrt{\left( E_x- E_y\right)^2 + \left(E_y-E_z\right)^2 + \left(E_z-E_x\right)^2},
\label{eqn:D}
\ee
which was previously studied in the compass model \cite{mishra:207201,wenzelQCMPRB,wenzelCM2010}. 
Here, $E_{x|y|z}$ is the total bond-energy along the ${x|y|z}$-direction. 

\paragraph{Critical exponents in the $e_g$ model ---} 
We start by presenting numerical results for the EgM~\eqref{eqn:eg_model} with SBC by displaying in Fig.~\ref{fig:overview}
the data for the magnetization $m$ and the Binder parameter $B_m=1-\langle m^4\rangle/ 3 \langle m^2 \rangle^2$ 
as a function of temperature.
Both observables indicate a continuous phase transition at about $T_c\approx 0.677$, in agreement with earlier PBC estimates~\cite{tanaka:267204,Rynbach}.
At $T_c$ we expect $B_m(L)$  to possess only corrections to scaling $B_m(L)=B_m^\star + cL^{-\omega}$ with $\omega$
being the correction exponent. We find our best estimate for $T_c=0.6775(1)$ and an effective $\omega\approx 1.4$ with a large constant $c$.

\begin{figure}[t]
\includegraphics[width=0.8\linewidth]{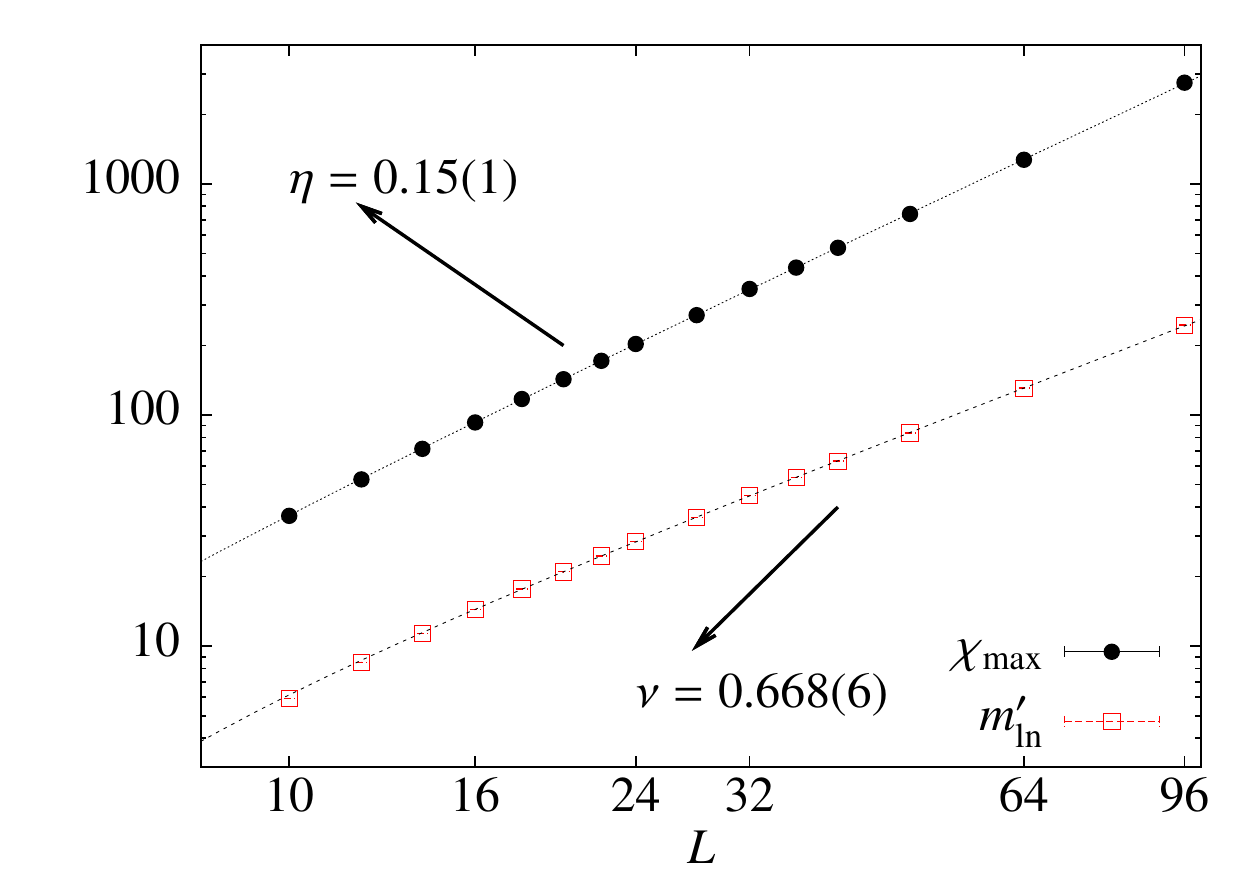}
\caption{\label{fig:FFS1}(Color online). {\em $e_g$ model}: Plot of $\chi_{\max}$ and $m'_{\ln}$ versus $L$ in  a double logarithmic scale. Estimates for $\nu$ and $\eta$ where obtained from a finite-size study using Eq.~\eqref{eq:ffs1}, taking into account corrections to scaling. The lines are the corresponding fit curves.}
\end{figure}
We now perform a finite-size scaling study to obtain the critical exponents. Here, we concentrate primarily on the correlation length 
exponent $\nu$ describing the divergence of the correlation length close to the critical point $\xi\sim \left|T-T_c\right|^{-\nu}$, as well 
as the exponent $\eta$ governing the decay of the spin-spin correlation function $G(r)\sim r^{-d+2-\eta}$ at the critical point.
We determine these exponents using the derivative of the logarithm of the order parameter 
$m'_{\ln}=\max\{(\mathrm{d} \ln m /\mathrm{d}\beta)\}$ \footnote{The slope of the Binder parameter gives consistent results but shows larger statistical fluctuations.} 
and the maximum of the susceptibility $\chi_{\max} = \max \{ N  \left( \langle m^2 \rangle - \langle m \rangle^2 \right) \}$ which are known to scale with 
system size $L$ as:~\cite{JankeMCGreifswald}
\begin{equation}
\label{eq:ffs1}
m'_{\ln}  \sim L^{1/\nu}(1+c_{m'} L^{-\omega}),\,\, \chi_{\max} \sim L^{2-\eta}(1+c_{\chi} L^{-\omega}).
\end{equation}
Using the effective correction exponent $\omega$ obtained above based
on the Binder cumulant, the data fits very well to Eq.~\eqref{eq:ffs1}
yielding our estimate $\nu=0.668(6)$ for the correlation length
exponent, see Fig.~\ref{fig:FFS1}. This value for $\nu$ would be
roughly consistent with the universality class of the 3D XY
universality) with $\nu_\mathrm{XY}=0.671$ \cite{CampostriniXY}.
However, an analogous analysis of the order parameter correlations at
criticality - from which we obtain $\eta=0.15(1)$ \cite{Footnote_eta}
- provides strong evidence for a universality class {\em distinct}
from the 3D XY class, which would yield a substantially smaller
$\eta_\mathrm{XY}
\approx0.038$~\cite{vicarireview,CampostriniXY}. Finally, an analysis
of the exponent $\alpha$ gives $\alpha\approx 0$ in agreement with the
usual hyper-scaling relation.

\paragraph{Critical exponents in the $e_g$-clock model---} 
\begin{figure}
\centering
\includegraphics[width=0.9\linewidth]{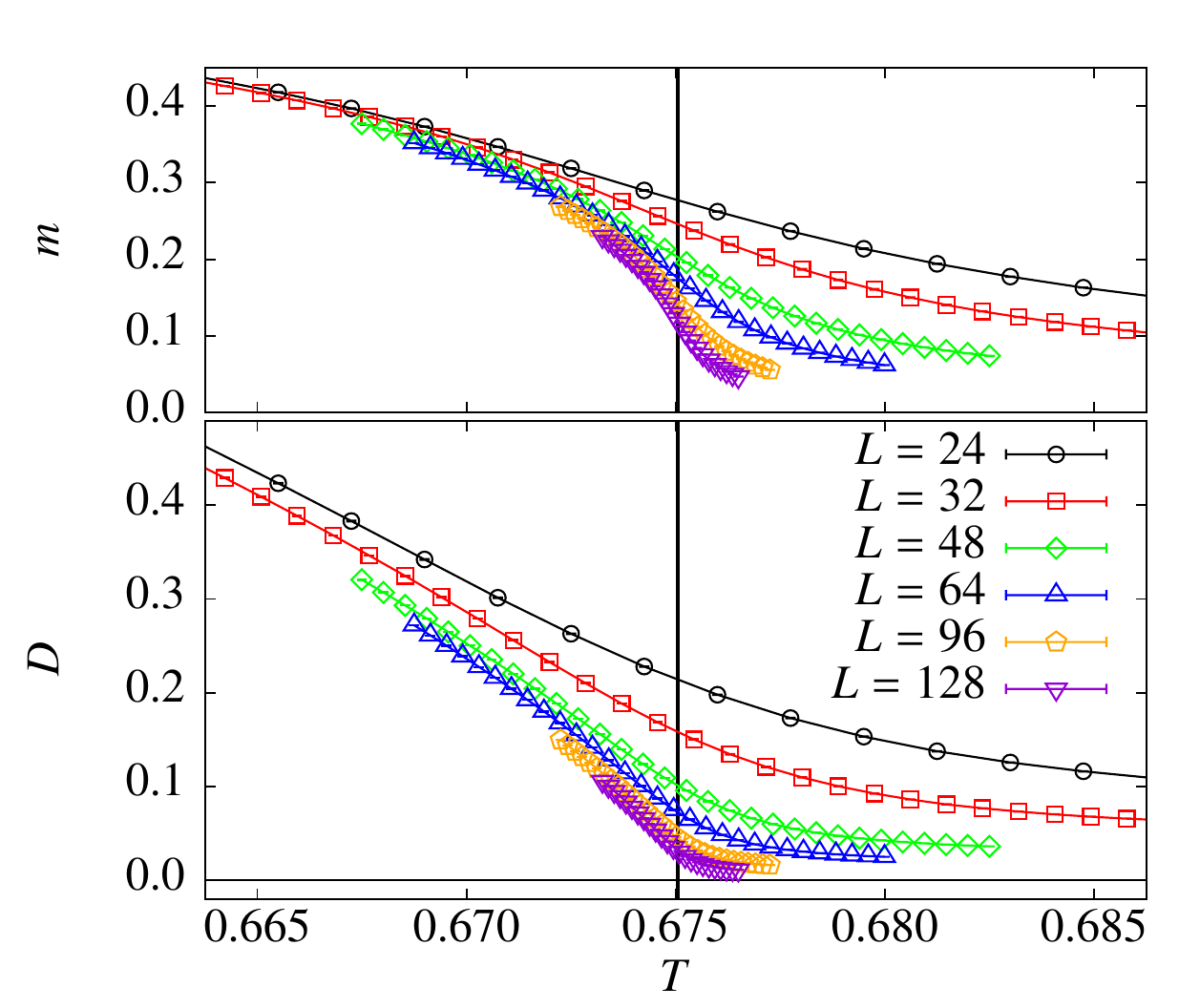}
\caption{\label{fig:EgCLM:D}(Color online). {\em $e_g$-clock model}: Orbital order parameter $m(T)$ (upper panel) and directional order parameter $D(T)$ (lower panel)
for different linear system sizes $L$. Note that both order parameters become finite below a common $T_c$ (indicated by the vertical line).}
\end{figure}
To investigate whether the continuous nature of the microscopic  degrees of freedom $\mathbf{T}$ has an impact on the critical properties,
we now consider a discrete version of Hamiltonian~\eqref{eqn:eg_model} -- one in which the vectors $\mathbf{T}$ can
only point along the six $\mathbb{T}^o_n$ ordering directions introduced above:
\begin{equation}
\label{egclock:model}
\mathcal{H}_{e_g}^\circledast= -\ J  \sum_{i,\alpha}  E^{\alpha}(n_i, n_{i+\mathbf{e}_\alpha})\,.
\end{equation}
Here, $E^{\alpha}(n_i,n_j)$ is the bond energy matrix along the bond direction $\alpha$ and $n = 0,\ldots, 5$ denote the six discrete onsite states. 
The similarity of our model to the 6-state ($Z_6$) clock model $\mathcal{H}_{Z_6}=- J \sum_{\langle i,j \rangle}  \mathbb{T}^o_{n_i}\cdot\mathbb{T}^o_{n_j}$~\cite{Potts1952},
suggests to term $\mathcal{H}_{e_g}^\circledast$  the \emph{$e_g$-clock model} (EgCLM). Its discrete nature allows to study larger systems of up to $L=128$.
In addition, we analyze the directional order parameter $D$ as introduced in Eq.~\eqref{eqn:D}. In an orbitally
ordered state characterized by a finite $m$, $D$ is also finite, however the converse is not true.
An illustrative example is given by the 2D compass model, where a gauge-like freedom forbids orbital ordering altogether~\cite{nussinovpip}, while
$D$ orders at finite temperature~\cite{mishra:207201,wenzelQCMPRB,wenzelCM2010}.

In Fig.~\ref{fig:EgCLM:D} we present data for $m(T)$ (upper panel) and $D(T)$ (lower panel) for different system sizes. Both $m$ and $D$ appear to set in 
at about the same temperature.
In order to confirm the simultaneous onset we have determined the respective Binder 
parameters $B_m$ and $B_D$ (not shown), 
indicating that both transitions take place at a unique critical temperature $T_c=0.67505(3)$.
This result rules out a scenario of a directionally ordered, orbital-disordered intermediate phase, and establishes a single transition from a high temperature disordered
phase to a low temperature orbitally ordered phase.

\begin{figure}
\centering
\includegraphics[width=0.9\linewidth]{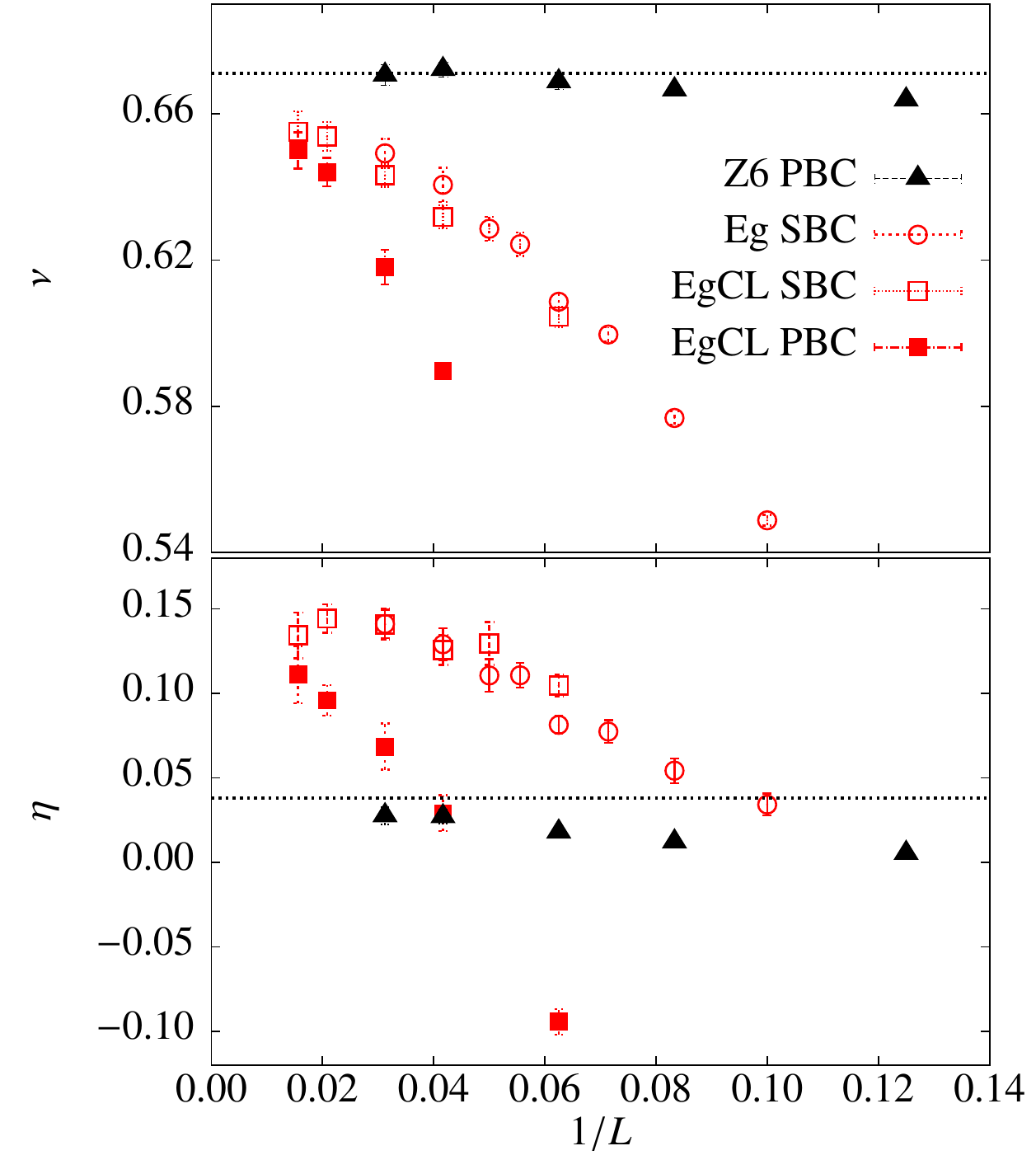}
\caption{\label{fig:ffs:exponents}
  (Color online). Finite-size scaling of the running exponents $\nu_L$ and $\eta_L$ calculated from Eqs.~\eqref{eq:nuL} and \eqref{eq:etaL} for several models and boundary conditions, see legend. 
  Both the $e_g$ and the $e_g$-clock model display the same critical behavior, which is different from the 3D XY universality class (indicated by the dashed lines and similar data 
  for the $Z_6$-clock model known to approach 3D XY universality~\cite{Jose_PRB}).}
\end{figure}
\begin{figure}[t]
\includegraphics[width=\columnwidth]{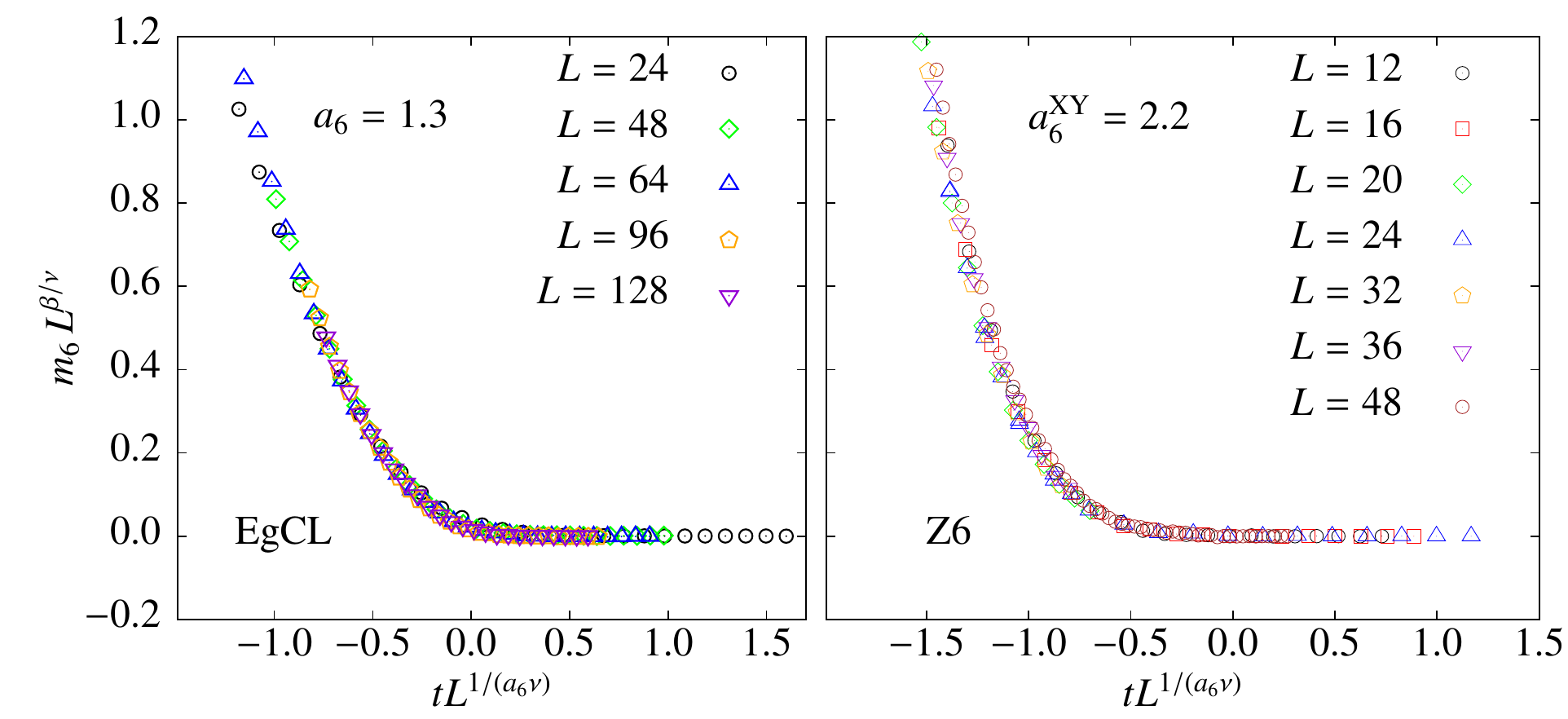}
\caption{(Color online). Collapse analysis of $m_6$ [see
  Eq.~\eqref{eqn:m6}] for the $e_g$-clock model (left) and the
  $Z_6$-clock model (right), based on the scaling assumption $m_6 \sim
  L^{-\beta/\nu}g(t L^{1/(a_6\nu)})$ (see Ref.~\cite{Louclock}). Best
  collapse parameters $a_6$ are indicated in the plot and differ clearly
  for the two models.
\label{fig:angulardependence}}
\end{figure}

Having demonstrated the simultaneity of the two ordering phenomena, we now perform a systematic study of the critical exponents in the EgCLM. 
Instead of fitting to Eq.~\eqref{eq:ffs1}, we study the finite-size behavior of (running) critical exponents obtained on system sizes $L$ and $2L$ via the relations 
\begin{align}
\label{eq:nuL}
\nu{_{L}} &= \ln(2)/\ln\left(m'_{\ln}(2L)/m'_{\ln}(L)\right),\\
\label{eq:etaL}
\eta{_{L}}  &= 2-\ln\left(\chi_{\max}(2L)/\chi_{\max}(L)\right)/\ln(2).
\end{align}
This allows to visualize finite-size effects directly and should give
the true exponents for $L\to \infty$.
In Fig.~\ref{fig:ffs:exponents} we present results for $\nu_L$ (upper panel) and $\eta_L$ (lower panel). 
In both quantities strong finite-size corrections are evident for the EgM and EgCLM, but our results convincingly show
that different boundary conditions (PBC/SBC) and both the EgM and the EgCLM converge to a single set of exponents: $\nu\approx 0.66$ 
and $\eta\approx 0.15$. These exponents - especially $\eta$ - are at variance with the corresponding values of the 3D XY
universality class. For comparison, we include data for the $Z_6$-clock model in Fig.~\ref{fig:ffs:exponents}, which quickly converges to the 
3D XY exponents expected for this model~\cite{Jose_PRB}.
Note that a similar analysis based on the order parameter $D$ instead of $m$ leads to the same $\nu$ exponent, while the corresponding $\eta_D$ 
exponent is much larger $(\approx 1.4)$. This simply follows from the assumption that $D$ has no intrinsic critical behavior, because then
$D$ is driven by $m$: $D \sim m^2$, resulting in an apparently different $\eta$ value.\\
\paragraph{Emergent U(1) symmetry ---} 
In order to shed light on the possible emergence of a U(1) symmetry at the critical point and the associated 
behavior of the crossover length scale $\Lambda$ for $T<T_c$ (as discussed in the context of $Z_q$-perturbed XY 
models~\cite{Jose_PRB,Blankschtein_PRB,Oshikawa_Zn,Louclock}), we determine the 6-fold anisotropy
$m_6$ of the orbital order $m$, based on order parameter histograms $P(r,\theta)$~\cite{Louclock}:
\be
\label{eqn:m6}
m_6= \int_0^1 d r \int_0^{2\pi}d\theta r^2 P(r,\theta) \cos(6\theta).
\ee
An analysis for the EgCLM analogous to Ref.~\cite{Louclock} yields a scaling of the crossover length $\Lambda$ with the correlation length $\xi$ as
$\Lambda \sim \xi^{a_6}$, with $a_6\approx 1.3$ [c.f. Fig.~\ref{fig:angulardependence}(a)]. In the case of a $Z_6$-perturbed 3D XY model 
we find $a_6^{XY}\approx 2.2$ (c.f.~Fig.~\ref{fig:angulardependence}(b), compatible with Ref.~\cite{Louclock}), almost a factor two larger than the value we obtain for the EgCLM.
%
\paragraph{Compass ($t_{2g}$) model ---} Finally we report our results for the second orbital-only model of interest here,
the $t_{2g}$ model in three dimensions~\cite{vandenBrink_NJP}, defined as:
\begin{equation}
\label{eqn:t2g_model}
    \mathcal{H}_{t_{2g}}= -\ J\sum_{i, \mathbf{e}_\alpha } \mathbf{T}_i^\alpha \mathbf{T}_{i+\mathbf{e}_\alpha}^\alpha,
\end{equation}
where now the degree of freedom $\mathbf{T}=(T^x,T^y,T^z)$ is a unit vector on the sphere $S^2$, and otherwise the notation follows Eq.~\eqref{eqn:eg_model}.
This model is also called 3D compass model
and is a straightforward generalization of the 2D compass model studied e.g.~in~\cite{mishra:207201,wenzelQCMPRB,wenzelCM2010}. 
An important difference of the $t_{2g}$ model compared to the $e_{g}$ model is that orbital order is ruled out due to the presence of 
gauge-like symmetries~\cite{nussinovpip}. Therefore, the order parameter $D$ [Eq.~\eqref{eqn:D}] can exhibit a phase transition in the 
absence of orbital ordering. We have simulated the full classical $t_{2g}$ model using the same simulation technology as for the $e_g$
model, revealing a {\em first order} transition at $T_c\approx 0.098$  from a high-temperature disordered to a low-temperature 
lattice symmetry broken phase indicated by a finite value of $D$. 

Recently the quantum $t_{2g}$ model has been studied using series expansions~\cite{Oitmaa2010}, and the absence of a phase transition at finite
temperature was conjectured. Our findings for the classical $t_{2g}$ model provide an alternative explanation as to why no (second order) finite-temperature transition
was detected: due to the first order nature, the transition is intrinsically difficult to detect based on series expansions. 
A detailed analysis of the properties of the $t_{2g}$ model will be presented in a forthcoming publication~\cite{WenzelEGlong}.

\paragraph{Conclusions ---} We have provided a detailed analysis of the critical properties of the finite-temperature ordering transitions in $e_g$ and $t_{2g}$ orbital-only models. While the $t_{2g}$ model
exhibits a first order transition, the critical properties of the
$e_g$ model point towards a distinct universality class, different
from the standard classes we have encountered so far. Further
theoretical work will be required to shed light on this observation,
and to understand in more detail the peculiar effects of the coupling
of real space and order parameter
space~\cite{vicarireview,Nattermann1975}, which are at work in these
models.
Given the broad range of systems where models similar to the ones studied 
here could arise (orbital systems in solids~\cite{vandenBrink_NJP,Jackeli2009}, Josephson junction arrays~\cite{gladchenko-2008}, and artificially engineered systems in optical lattices~\cite{Duan2003}),
we are optimistic that the peculiar critical properties uncovered in the present work can be further explored experimentally.

\acknowledgments
We thank M.~Hasenbusch, G.~Misguich, R.~Moessner, M.~Oshikawa, and S.~Trebst for useful discussions. 
The simulations have been performed on the PKS-AIMS cluster at the MPG RZ Garching
and on the Callisto cluster at EPF Lausanne.


\begin{thebibliography}{10}

\bibitem{vandenBrink_NJP}
J.~van~den Brink,
\href{http:dx.doi.org/10.1088/1367-2630/6/1/201}{New J. Phys. {\bf 6}, 201 (2004)}.

\bibitem{kitaev-model}
A.Y.~Kitaev,
\href{http:dx.doi.org/10.1016/j.aop.2005.10.005}{Ann. Phys. {\bf 321}, 2 (2006)}.

\bibitem{kugel82}
K.I.~Kugel and D.I.~Khomskii,
\href{http:dx.doi.org/10.1070/PU1982v025n04ABEH004537}{Sov. Phys. Usp. {\bf 25}, 231 (1982)}.

\bibitem{doucot:024505}
B.~Dou\c{c}ot {\it et al.},
\href{http:dx.doi.org/10.1103/PhysRevB.71.024505}{Phys. Rev. B {\bf 71}, 024505 (2005)}.

\bibitem{gladchenko-2008}
S.~Gladchenko {\it et al.},
\href{http:dx.doi.org/10.1038/nphys1151}{Nat. Phys. {\bf 5}, 48 (2009)}.

\bibitem{vicarireview}
A.~Pelissetto and E.~Vicari,
\href{http:dx.doi.org/10.1016/S0370-1573(02)00219-3}{Phys. Rep. {\bf 368}, 549 (2002)}.

\bibitem{mishra:207201}
A.~Mishra {\it et al.},
\href{http:dx.doi.org/10.1103/PhysRevLett.93.207201}{Phys. Rev.~Lett. {\bf 93}, 207201 (2004)}.

\bibitem{wenzelQCMPRB}
S.~Wenzel and W.~Janke,
\href{http:dx.doi.org/10.1103/PhysRevB.78.064402}{Phys. Rev. B {\bf 78}, 064402 (2008)}.

\bibitem{wenzelCM2010}
S.~{Wenzel}, W.~{Janke}, and A.~M. {L{\"a}uchli},
\href{http:dx.doi.org/10.1103/PhysRevE.81.066702}{Phys. Rev. E {\bf 81}, 066702 (2010)}.

\bibitem{Rynbach}
A.~van~Rynbach, S.~Todo, and S.~Trebst,
\href{http:dx.doi.org/10.1103/PhysRevLett.105.146402}{Phys. Rev. Lett. {\bf 105}, 146402 (2010)}.

\bibitem{nussinov-2004-6}
Z.~Nussinov {\it et al.},
\href{http:dx.doi.org/10.1209/epl/i2004-10134-5}{Europhys. Lett. {\bf 67}, 990 (2004)};
M.~Biskup, L.~Chayes, and Z.~Nussinov,
\href{http:dx.doi.org/10.1007/s00220-004-1272-7}{Commun. Math. Phys. {\bf 255}, 253 (2005)}.

\bibitem{tanaka:267204}
T.~Tanaka, M.~Matsumoto, and S.~Ishihara,
\href{http:dx.doi.org/10.1103/PhysRevLett.95.267204}{Phys.~Rev.~Lett. {\bf 95}, 267204 (2005)}.

\bibitem{Jose_PRB}
J.V.~Jos\'e {\it et al.},
\href{http:dx.doi.org/10.1103/PhysRevB.16.1217}{Phys. Rev. B {\bf 16}, 1217 (1977)}.

\bibitem{Blankschtein_PRB}
D.~Blankschtein {\it et al.},
\href{http:dx.doi.org/10.1103/PhysRevB.29.5250}{Phys. Rev. B {\bf 29}, 5250 (1984)}.

\bibitem{Oshikawa_Zn}
M.~Oshikawa,
\href{http:dx.doi.org/10.1103/PhysRevB.61.3430}{Phys. Rev. B {\bf 61}, 3430 (2000)}.

\bibitem{Louclock}
J.~Lou, A.W.~Sandvik, and L.~Balents,
\href{http:dx.doi.org/10.1103/PhysRevLett.99.207203}{Phys. Rev. Lett. {\bf 99}, 207203 (2007)}.

\bibitem{FAlet-dimer}
F.~Alet {\it et al.},
\href{http:dx.doi.org/10.1103/PhysRevLett.97.030403}{Phys. Rev. Lett. {\bf 97}, 030403  (2006)}.

\bibitem{CharrierAlet2010}
D.~Charrier and F.~Alet,
\href{http:dx.doi.org/10.1103/PhysRevB.82.014429}{Phys. Rev. B {\bf 82}, 014429 (2010)}.

\bibitem{WenzelEGlong}
S.~Wenzel and A.~M. L\"auchli.
unpublished (2011).

\bibitem{Note1}
The slope of the Binder parameter gives consistent results but shows larger
statistical fluctuations.

\bibitem{JankeMCGreifswald}
W.~Janke,
\href{http:dx.doi.org/10.1007/978-3-540-74686-7_4}{Lect. Notes Phys. {\bf 739}, 79 (2008)}.

\bibitem{CampostriniXY}
M.~Campostrini {\it et al.},
\href{http:dx.doi.org/10.1103/PhysRevB.63.214503}{Phys. Rev. B {\bf 63}, 214503 (2001)}.

\bibitem{Footnote_eta}
A straight line fit for $L>24$ yields $\eta\approx 0.13$.

\bibitem{Potts1952}
R.~B. Potts,
\href{http:dx.doi.org/10.1017/S0305004100027419}{Proc. Camb. Philos. Soc. {\bf 48}, 106 (1952)}.

\bibitem{nussinovpip}
Z.~Nussinov and E.~Fradkin,
\href{http:dx.doi.org/10.1103/PhysRevB.71.195120}{Phys. Rev. B {\bf 71}, 195120 (2005)}.

\bibitem{Oitmaa2010}
J.~Oitmaa and C.J.~Hamer,
\href{http:dx.doi.org/10.1103/PhysRevB.83.094437}{Phys. Rev. B {\bf 83}, 094437 (2011)}.

\bibitem{Nattermann1975}
T.~Nattermann and S.~Trimper,
\href{http:dx.doi.org/10.1088/0305-4470/8/12/016}{J. Phys. A: Math. Gen. {\bf 8}, 2000 (1975)}.

\bibitem{Jackeli2009}
G.~Jackeli and G.~Khaliullin,
\href{http:dx.doi.org/10.1103/PhysRevLett.102.017205}{Phys. Rev. Lett. {\bf 102}, 017205 (2009)};
J.~Chaloupka, G.~Jackeli, and G.~Khaliullin,
\href{http:dx.doi.org/10.1103/PhysRevLett.105.027204}{Phys. Rev. Lett. {\bf 105}, 027204 (2010)}.

\bibitem{Duan2003}
L.-M. Duan, E.~Demler, and M.~D. Lukin,
\href{http:dx.doi.org/10.1103/PhysRevLett.91.090402}{Phys. Rev. Lett. {\bf 91}, 090402 (2003)};
A.~Micheli, G.K. Brennen, and P.~Zoller,
\href{http:dx.doi.org/10.1038/nphys287}{Nat. Phys. {\bf 2}, 341 (2006)}.

\end{thebibliography}
\end{document}